\documentclass[aps,prl,reprint,superscriptaddress]{revtex4-2}
\usepackage{graphicx}
\usepackage{amsmath}% 
\usepackage{color}
\usepackage{booktabs}
\usepackage{makecell}
\usepackage{multirow}
\usepackage{textcomp}
\usepackage{setspace}
\begin{document}

\title{Towards understanding the defect properties in the multivalent A-site Na$_{0.5}$Bi$_{0.5}$TiO$_3$-based perovskite ceramics}  
 
\author{Pengcheng Hu} 
%\altaffiliation[These authors contributed equally]{}
\affiliation{Institute of Materials Science, Electronic Structure of Materials, Technical University of Darmstadt, Otto-Berndt-Str 3, Darmstadt 64287, Germany}

\author{Chinmay Chandan Parhi} 
\affiliation{Institute for Chemistry and Technology of Materials, Graz University of Technology, Stremayrgasse 9, Graz 8010, Austria}

\author{Jurij Koruza}
\affiliation{Institute for Chemistry and Technology of Materials, Graz University of Technology, Stremayrgasse 9, Graz 8010, Austria}

\author{Andreas Klein}
\affiliation{Institute of Materials Science, Electronic Structure of Materials, Technical University of Darmstadt, Otto-Berndt-Str 3, Darmstadt 64287, Germany}
\email{aklein@esm.tu-darmstadt.de}

\date{\today}

\begin{abstract} 

A defect model involving cation and anion vacancies and anti-site defects is proposed that accounts for the non-stoichiometry of multi-valent $A$-site Na$_{0.5}$Bi$_{0.5}$TiO$_3$ based perovskite oxides with $AB$O$_3$ composition. A series of samples with varying $A$-site non-stoichiometry and $A$:$B$ ratios were prepared to investigate their electrical conductivity. The oxygen partial pressure and temperature dependent conductivities where studied with direct current (dc) and alternating current (ac) techniques, enabling to separate between ionic and electronic conduction. The Na-excess samples, regardless of the $A$:$B$ ratio, exhibit dominant ionic conductivity and $p$-type electronic conduction, with the highest total conductivity reaching $4 \times 10^{-4}$\,S/cm at 450$^\circ$C. In contrast, the Bi-excess samples display more insulating characteristics and $n$-type electronic conductivity, with conductivity values within the 10$^{-8}$ S/cm range at 450$^\circ$C. These conductivity results strongly support the proposed defect model, which offers a straightforward description of defect chemistry in NBT-based ceramics and serves as a valuable guide for optimizing sample processing to achieve tailored properties.

\end{abstract}
\maketitle 
\section{Introduction} 

The electrical conductivity of oxide ceramics can be broadly categorized into electronic and (oxygen) ionic conductivity \cite{kharton2004transport}. Based on the oxide-ion transport number ($t_{\mathrm{ion}}$), the conductivity of a material can be divided into three distinct types: Type I ($t_{\mathrm{ion}}<$ 0.1): predominantly electronic conductivity, as observed in materials like Sn-doped In$_2$O$_3$ (ITO) \cite{kostlin1975optical}, and LaCrO$_3$ \cite{sakai1990sinterability}; Type II ($t_{\mathrm{ion}}>$ 0.9): predominantly ionic conductivity, exemplified by materials such as Y-stabilized ZrO$_2$ (YSZ) \cite{badwal1984electrical}; Type III (0.1 $<t_{\mathrm{ion}}$ $<$ 0.9): mixed conductivity, where both electronic and ionic conduction mechanisms play a significant role, as seen in LaCoO$_3$ (LCO) \cite{skinner2003oxygen} or (La,Sr)(Co,Fe)O$_3$ (LSCO) \cite{tuller2017ionic}.

In order to distinguish the contribution of ionic and electronic conduction, impedance spectroscopy, electromotive force (EMF), and $^{18}$O Time-of-Flight Secondary Ion Mass Spectrometry (ToF-SIMS) are commonly employed to determine the ionic transport number ($t_{\textsubscript{ion}}$) \cite{li2015dramatic, li2014family}. In this work, we test a simpler approach to discriminate between ionic and electronic conduction by comparing direct current (dc) and alternating current (ac) conductivity. The dc method probes electronic conduction, as ionic transport is blocked by the electrodes (Platinum) at not too high temperature \cite{baiatu1990dc}. In contrast, ac conductivity reflects the total conductivity, probing all mobile charge carriers within the sample through the application of an alternating signal \cite{macdonald2018impedance}. 

Then the ionic conduction can be determined according to:

\begin{equation}
 \sigma_{\mathrm{ionic}} = \sigma_{\mathrm{total}} - \sigma_{\mathrm{electronic}} = \sigma_{\mathrm{ac}}-\sigma_{\mathrm{dc}}\\
\end{equation}

Sodium bismuth titanate (Na$_{0.5}$Bi$_{0.5}$TiO$_3$, NBT) was first reported by Smolenskii in 1961 \cite{smolenskii1961new}. It is a perovskite-type compound in which the $A$-site is formally occupied by equal amounts of monovalent Na and trivalent Bi, while the $B$-site is occupied by tetravalent Ti. Of particular interest are NBT-based solid solutions, like NBT-BaTiO$_3$ \cite{takenaka1991bi1}, NBT-SrTiO$_3$ \cite{koruza2016formation}, or NBT-BaTiO$_3$-(K,Na)NbO$_3$ \cite{zhang2007giant}, which enable high electromechanical strains due to field-induced phase transitions \cite{paterson2018relaxor} and good high-power performance \cite{slabki2022origin}. NBT and NBT-BaTiO$_3$, the more application-relevant compositions, where also Ba is occupying the A-sites can exhibit all three types of conductivity behaviors by altering its $A$-site stoichiometry, enabling it to serve diverse applications \cite{yang2018defect,ren2024enhanced,steiner2019effect}. On one hand, nominally stoichiometric, Bi-excess on the $A$-site, or donor-doped samples are considered among the most promising lead-free piezoelectric materials for replacing lead zirconate titanate (PZT). These compositions suppress oxide-ion conductivity in NBT, enhancing its insulation characteristics and reducing the risk of dielectric breakdown during the poling process of ceramics \cite{takenaka1991bi1,slabki2022origin,li2015donor}. On the other hand, samples with Na-excess on the $A$-site and acceptor-doped samples (e.g., Mg, Fe, or Zn) exhibit ionic conductivities that make them  candidate materials for electrolytes in intermediate-temperature solid oxide fuel cells (SOFCs) \cite{yang2017optimisation,singhproperties,zhang2011ab}. The high concentration and mobility of oxygen vacancies, induced during material processing, facilitates ionic conductivity of up to 0.01 S/cm at $600^\circ$C, comparable with that of conventional oxide ion conductors like YSZ \cite{li2014family,steiner2019effect,yang2018defect}. 

Currently, no comprehensive defect model exists for mixed-valence perovskites in which either the $A$-site or the $B$-site is occupied by ions of differing valences. Notable examples of such materials include NBT and lead magnesium niobate (PbMg$_{1/3}$Nb$_{2/3}$O$_3$, PMN). Existing analyses of defect equilibria in such compounds often rely on restricted equilibria, which do not account for the full range of possible defects. Such oversight can result in misinterpretation of defect behavior in these systems. A fundamental challenge in modeling defects in mixed-valence perovskites is the presence of varying cation-ordering configurations (see e.g.\ \cite{Groeting11}, which disables a unique assignment of specific lattice sites). For instance, when a cation is removed from the $A$-site, the resulting vacancy could correspond to either Na or Bi, yet determining its exact origin remains unclear as there are no uniquely defined lattice positions. Apart from this ambiguity, cation vacancies generally behave as acceptor defects, typically compensated by either free or trapped holes, or oxygen vacancies \cite{Klein2023}. 

In this work, a defect model will be introduced to predict the dominant conduction mechanism based on the determining compositional parameters, the Na:Bi and the $A$:$B$ ratios, as well as doping effects. The results of the defect model will be compared to electrical conductivity measurements using 0.94(Na$_{0.50}$Bi$_{0.50}$)TiO$_3$-0.06BaTiO$_3$ (NBT-6BT) with various composition, demonstrating that the model is able to predict the nature of conductivity of NBT-based ceramics. 

\section{Defect model} 
\label{Defect model}

For simplicity, we assume that no secondary phases are present. This assumption may not be entirely accurate as variations in the $A$:$B$ and Na:Bi ratios may directly result in the formation of secondary phases. The likelihood of this occurrence depends on the width of the stability region and the extent to which the Fermi energy is influenced by off-stoichiometry. A comprehensive analysis of this phenomenon in BaTiO$_3$ was conducted by Lee et al.\ \cite{lee2007influence, lee2007modified, lee2008comprehensive, lee2008comprehensive1}, who demonstrated that the stability window for the Ba:Ti ratio lies within 0.9817 $\le$ Ba:Ti $\le$ 1.0108 when sintered at 1400$^{\circ}$C. The situation, however, will be different for other compounds.

Relevant defects, which can, in principle, be present in NBT and in NBT-BT are outlined in Table \ref{species}. The listing ignores for the moment the lack of a unique assignment of $A$-lattice sites to either Na or Bi. The possibility for lower charge states, for example a singly charged or neutral oxygen vacancies, is also not included. The substitutional defect species arise from the solid solution of NBT with BaTiO$_3$, which is widely used for piezoelectric applications. In this case, Ba may  substitute either for Na or Bi, constituting either a donor, $\mathrm{Ba_{Na}^{\cdot}}$, or an acceptor, $\mathrm{Ba_{Bi}^{\prime}}$. The substitutional defects cannot be ignored, as the formation energy for donors and acceptor species have an opposite dependence on the Fermi level \cite{Klein2023}. Depending on the other defect concentrations, this may result in a preference for substituting either of the $A$-site species and hence affect the Na:Bi ratio. This complication will not be treated explicitly in this contribution. So far, there is also no clear evidence indicating an influence of Ba on the defect equilibria.  

\begin{table}[ht]
	\centering
	\caption{Potentially relevant defect species in NBT-BT ceramics.}
	\label{species}
	\vspace{5pt}
	\small
	\setlength{\tabcolsep}{2.5mm}{
		\begin{tabular}{lcc}
			\hline \hline
			%\multicolumn{2}{c}{Negatively charged point defects}  \\ 	\hline 
            defect & negative & positive \\ \hline 
			free electron/hole & e$^{\prime}$ & h$^{\cdot}$  \\
			trapped electron/hole & e$_{\mathrm{p}}^{\prime}$ & h$_{\mathrm{p}}^{\cdot}$  \\
            antisite & $\mathrm{Na_{Bi}^{\prime\prime}}$ & $\mathrm{Bi_{Na}^{\cdot\cdot}}$ \\
		      vacancies & ${V_\mathrm{Na}^{\prime}}$, ${V_\mathrm{Bi}^{\prime\prime\prime}}$, ${V_\mathrm{Ti}^{\prime\prime\prime\prime}}$ &  ${V_\mathrm{O}^{\cdot\cdot}}$ \\ 
            substitution& $\mathrm{Ba_{Bi}^{\prime}}$ & $\mathrm{Ba_{Na}^{\cdot}}$ \\ \hline 
			\end{tabular}}	
\end{table} 

To account for the fundamental impossibility of assigning unique lattice sites to the $A$-site species Na and Bi and to reduce the number of contributing defect species, we start by defining two parameters, which are sufficient to completely determine the cation stoichiometry:
\begin{equation}
	\begin{split}
		X&= [\mathrm{Na}_A]:[\mathrm{Bi}_A]  \\
		Y&= [A_A]:[B_B]  \\ 
	\end{split} 
\end{equation}
where $[\ldots]$ represent site fractions, $X$ and $Y$ should not deviate much from unity, and 
\begin{equation}
    {[A_A]} = [\mathrm{Na}_A] + [\mathrm{Bi}_A]
\end{equation}

For the $A$-site elements, it is assumed that the number of Na and Bi sites is formally equal. Rather than specifying their specific positions, we define them solely by their quantities. Additionally, the effects of $A$-site non-stoichiometry are formally considered through the formation of $\mathrm{Na_{Bi}^{\prime\prime}}$ and $\mathrm{Bi_{Na}^{\cdot\cdot}}$ antisite defects, while variations in the $A$:$B$ ratio influence the formation of $A$-site or $B$-site cation vacancies. The relevant defect concentrations in NBT can then be expressed with the help of $X$ and $Y$ as:

\begin{equation}
	\label{equation1}
	\begin{split}
		\mathrm{[Na_{Bi}^{\prime\prime}]} &= \frac{1}{2}\cdot \frac{X-1}{X+1} \\
		\mathrm{[Bi_{Na}^{\cdot\cdot}]} &= -\frac{1}{2}\cdot \frac{X-1}{X+1} \\
		{[V_A^{\prime\prime}]} &= 1-Y \quad \mathrm{for} \quad Y<1 \\
		{[V_B^{\prime\prime\prime\prime}]} ={[V_\mathrm{Ti}^{\prime\prime\prime\prime}]} &= Y-1 \quad \mathrm{for} \quad Y>1 %\\
		%\mathrm{[Ba_{Bi}^{\prime}]}&=xf(X,Y) \\
		%\mathrm{[Ba_{Na}^{\cdot}]}&=xf_{1}(X,Y)
	\end{split}
\end{equation}

Based on this framework, all defect concentrations can be quantitatively determined from the stoichiometry of the NBT samples. The effects of $A$-site non-stoichiometry will be discussed first:

\begin{itemize}
\item For $X=1$, the equations reveal:
\begin{equation}
    \begin{split}
		[\mathrm{Na}_{A}] &= [\mathrm{Bi}_{A}] = 0.5   \\ 
       \mathrm{and} \quad \mathrm{[Na_{Bi}^{\prime\prime}]} &= \mathrm{[Bi_{Na}^{\cdot\cdot}]} = 0	
    \end{split}
\end{equation}
As long as $\mathrm{[Na_{Bi}^{\prime\prime}]=[Bi_{Na}^{\cdot\cdot}]}$, their charges are balancing each other, allowing us to disregard these antisite defects. Antisite defects only become significant when $X$ deviates from 1.

\item For $X\ne 1$, $X$ can be treated as equivalent to doping, and the concentrations of the antisite defects for all cases of $X$ can be expressed as follows:
\begin{equation}
	\begin{split}
		X>1:& \quad \mathrm{[Na_{Bi}^{\prime\prime}]} = \frac{1}{2}\cdot \frac{X-1}{X+1} \quad \& \quad \mathrm{[Bi_{Na}^{\cdot\cdot}]=0} \\
		X<1:& \quad \mathrm{[Bi_{Na}^{\cdot\cdot}]} = -\frac{1}{2}\cdot \frac{X-1}{X+1} \quad \& \quad \mathrm{[Na_{Bi}^{\prime\prime}]=0} \\
	\end{split}
\end{equation}
\end{itemize}
A variation in the $A$:$B$ ratio is accomplished by the formation of either $A$-site or $B$-site vacancies. The concentrations of cation vacancies can then be expressed as:
\begin{itemize}
	\item $Y=1$, the concentrations follow from thermodynamic defect equilibrium. 
    
	\item $Y>1$, The material is $A$-site rich, leading to the formation of $B$-site vacancies to maintain the stoichiometric ratio of $A$ and $B$ lattice sites in the perovskite lattice. Then ${[A]:([B]+[V_B])=1}$ and the concentration of $B$-site (Ti) vacancies can be calculated as:
	\begin{equation}
		\mathrm{[V_\mathrm{Ti}^{\prime\prime\prime\prime}]} = Y-1 
	\end{equation}
    
	\item $Y<1$, the material is $B$-site rich, leading to the formation of $A$-site vacancies with a total concentration of ${[V_A^{\prime\prime}]}  =1-Y$. However, it is unclear which specific vacancies are formed. As the effect of the Na:Bi ratio is considered independently by the parameter $X$, the material should generate equal numbers of Na and Bi vacancies with an average charge of 2 per vacancy.
   \begin{equation}
		\begin{split}
			[V_\mathrm{Na}^{\prime}] = [V_\mathrm{Bi}^{\prime\prime\prime}] = \frac{[V_A]}{2} = \frac{1-Y}{2} \quad \\
			\text{average charge per vacancy = 2}\\ 
		\end{split}
	\end{equation}
\end{itemize}

Based on the model outlined above, the samples exhibit specific $A$:$B$ and Na:Bi ratios along with their corresponding defect concentrations. For the calculations, it is assumed that $A$-site vacancies are, on average, doubly charged. Samples with high conductivity at high temperature (e.g., 450$^\circ$C) are likely to be effectively acceptor-doped, compensated by a high concentration of oxygen vacancies. In contrast, samples with low conductivity at high temperature are effectively donor-doped, reflecting a low concentration of oxygen vacancies.

The effective doping effect can then be determined based on the composition. For example, for a sample containing 51\,\% Na and 49\,\% Bi on the $A$-site (51/49), $X=1.042$ and $Y=1$, while for a sample containing 49\,\% Na and 51\,\% Bi on the $A$-site (49/51), $X=0.98$ and $Y=1$. For the two examples, the effective doping concentrations are:
\begin{equation}
	\begin{split}
	\text{51/49:} \quad & [\mathrm{acc}] = 2 \cdot [\mathrm{Na}_\mathrm{Bi}^{\prime\prime}] + 2 \cdot [V_A^{\prime\prime}] + 4 \cdot [V_\mathrm{Ti}^{\prime\prime}] \\ 
    &\quad = (1.042-1)/(1.042+1) = 0.021  \\ 
	\text{49/51:} \quad & [\mathrm{don}] = 2 \cdot [\mathrm{Bi}_\mathrm{Na}^{\cdot\cdot}] - 2 \cdot [V_A^{\prime\prime}] - 4 \cdot [V_\mathrm{Ti}^{\prime\prime}] \\ 
    &\quad = (0.961-1)/(0.961+1) = 0.020 \\
    \end{split}
\end{equation}
If the $A$:$B$ ratio deviates from unity, additional acceptors are generated, which may override the donor concnetration of Bi-rich samples. 

\section{Experimental} 

\begin{figure}[ht]
	\centering
	\includegraphics[scale=0.9]{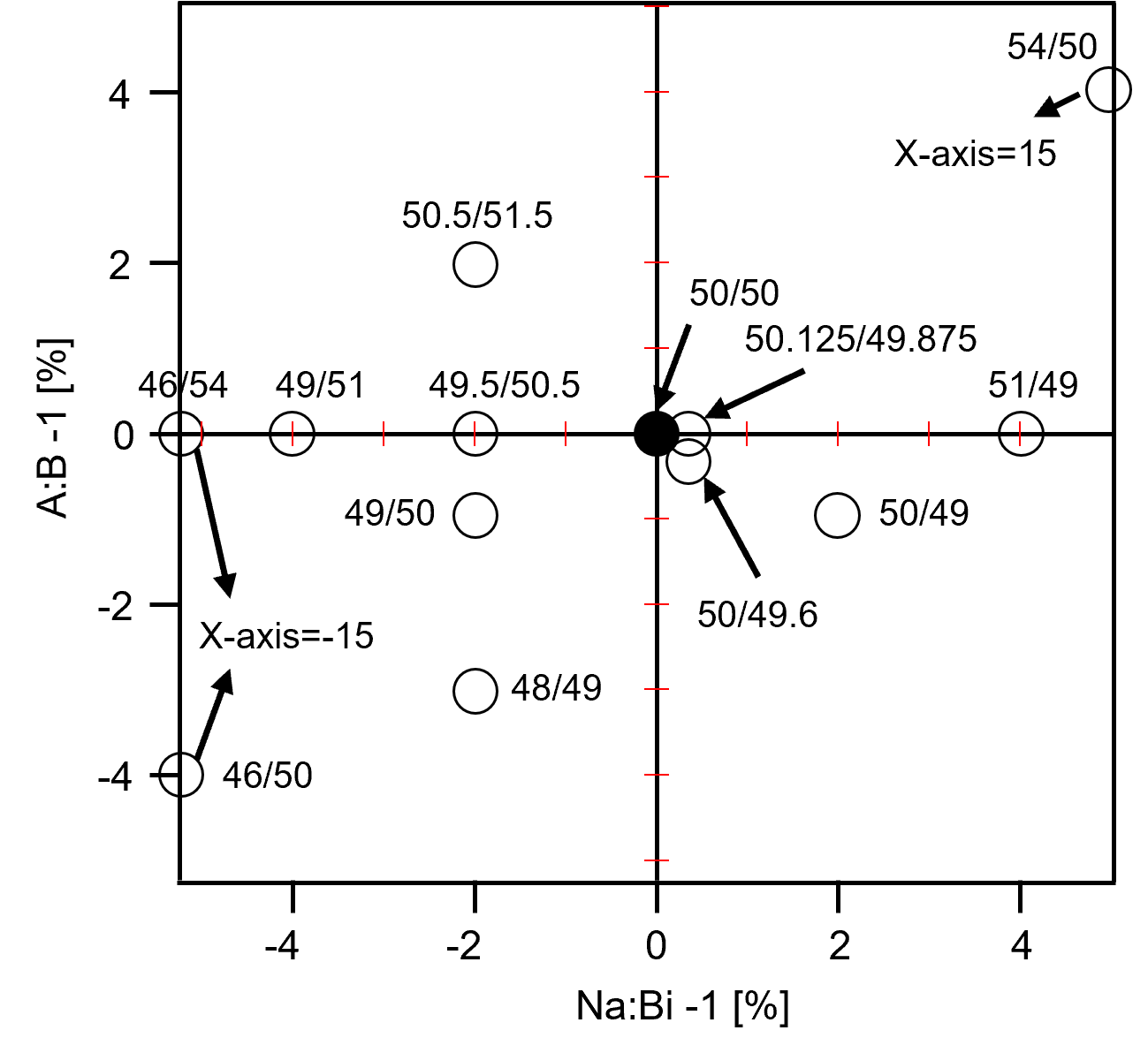}
	\centering
	\caption{Na:Bi and $A$:$B$ ratios of all the NBT-6BT samples used in this work.} 
	\label{fig sample}
\end{figure}

To verify the validity of the defect model, a series of 0.94(Na$_{0.5}$Bi$_{0.5}$)TiO$_3$-0.06BaTiO$_3$ (NBT-6BT) samples with varying Na:Bi- and $A$:$B$-ratios have been prepared by solid-state synthesis. Powders with given purities were utilized: Na$_2$CO$_3$ (99.95\,\%), BaCO$_3$ (99.95\,\%), Bi$_2$O$_3$ (99.975\,\%) (all from Alfa Aesar GmbH \& Co. KG, Germany) and TiO$_2$ (99.8\,\%, Sigma Aldrich). After weighing the precursors, the powders were milled with yttria-stabilized zirconia balls in ethanol for 4\,h at 250\,rpm, dried and homogenized, and finally calcined at 900$^\circ$C for 3 h using a heating rate of $5\,^\circ$C/min. Milling was subsequently repeated under the same conditions and was followed by one more drying step. Disk samples of 13\,mm in diameter were pressed at a uniaxial pressure of 15.4\,MPa followed by an isostatic pressure of 300\,MPa. The samples were placed in a closed alumina crucible with sacrificial powder and were sintered at $1150\,^\circ$C for 3\,h using a ramp rate of $5\,^\circ$C/min. The compositions chosen for the experiments are summarized in Table \ref{compo}. The Na:Bi- and $A$:$B$-ratios of all samples are furthermore illustrated in Fig. \ref{fig sample}.

\begin{table}[ht]
	\renewcommand\arraystretch{1.5}
	\centering
	\caption{Summary of all NBT-6BT ceramics used for the electrical conductivity measurements. }
	\label{compo}
	\vspace{5pt}
	\small
	\setlength{\tabcolsep}{1.0mm}{
		\begin{tabular}{lc}
			\hline \hline
			Bulk material &Abbreviation  \\ \hline 
			0.94(Na$_{0.50}$Bi$_{0.50}$)TiO$_3$-0.06BaTiO$_3$ &50/50 \\
		      0.94(Na$_{0.51}$Bi$_{0.49}$)TiO$_3$-0.06BaTiO$_3$ &51/49  \\
			  0.94(Na$_{0.46}$Bi$_{0.54}$)TiO$_3$-0.06BaTiO$_3$ &46/54  \\
			0.94(Na$_{0.49}$Bi$_{0.5}$)TiO$_3$-0.06BaTiO$_3$ &49/50 \\
		      0.94(Na$_{0.50}$Bi$_{0.49}$)TiO$_3$-0.06BaTiO$_3$ &50/49 \\
		      0.94(Na$_{0.54}$Bi$_{0.5}$)TiO$_3$-0.06BaTiO$_3$ &54/50 \\
		      0.94(Na$_{0.46}$Bi$_{0.5}$)TiO$_3$-0.06BaTiO$_3$ &46/50  \\
            0.94(Na$_{0.50}$Bi$_{0.496}$)TiO$_3$-0.06BaTiO$_3$ &50/49.6 \\
            0.94(Na$_{0.49}$Bi$_{0.51}$)TiO$_3$-0.06BaTiO$_3$ &49/51  \\
            0.94(Na$_{0.495}$Bi$_{0.505}$)TiO$_3$-0.06BaTiO$_3$ &49.5/50.5 \\
            0.94(Na$_{0.505}$Bi$_{0.515}$)TiO$_3$-0.06BaTiO$_3$ &50.5/51.5 \\
            0.94(Na$_{0.48}$Bi$_{0.49}$)TiO$_3$-0.06BaTiO$_3$ &48/49 \\
            0.94(Na$_{0.50125}$Bi$_{0.49875}$)TiO$_3$-0.06BaTiO$_3$ &50.125/49.875 \\
            1\% Zn-doped 0.94Na$_{0.50}$Bi$_{0.50}$)TiO$_3$-0.06BaTiO$_3$ &Zn-doped \\
			\hline \hline			
	\end{tabular}}	
\end{table}

For electrical conductivity measurements, the sintered samples were ground to a thickness of 0.35\,mm. Platinum electrodes were sputtered onto both sides for electrical measurements after annealing at $450\,^\circ$C for 1\,h to relieve the residual stresses introduced by grinding. %X-ray diffraction (XRD) was performed on powder samples, prepared by crushing and annealing the sintered pellets. A Rigaku-600 X-ray diffractometer and a Cu-source (Cu K$\alpha$1 $\lambda$ = 1.5409 $\AA$) is used in transmission geometry. 
The oxygen partial pressure dependence of conductivity was measured by systematically decreasing the oxygen partial pressure from $10^5$\,ppm to 1\,ppm in steps of one order of magnitude. The actual oxygen partial pressure  was set and monitored using a SGM5-EL electrolysis device (ZIROX Sensoren $\&$ Elektronik GmbH, Greifswald, Germany). Direct current 2-point conductivity measurements were performed at a constant dc voltage of 0.5\,V at $450\,^\circ$C, with the current recorded using a Keithley 6487 picoammeter (Tektronix, Inc., USA). Alternating current (ac) measurements (impedance spectroscopy) were performed by the novocontrol electrochemical impedance system (Novocontrol Technologies GmbH \& Co. KG, Montabaur, Germany) with the frequency range from $10^7 - 0.01$\,Hz. The temperature dependence of conductivity was measured by two methods: the \textit{static} ac method at constant temperature with an ac amplitude of 0.5\,V in dry air, and the \textit{dynamic} dc method during temperature ramping with a rate of 2.5\,K/min using a dc voltage of 0.5 V\,in dry air and in N$_2$. 

\section{Experimental Results and Discussion}

\subsection{Oxygen partial pressure dependent conductivity}

\begin{figure}
    \centering
    \includegraphics[width=8.5cm]{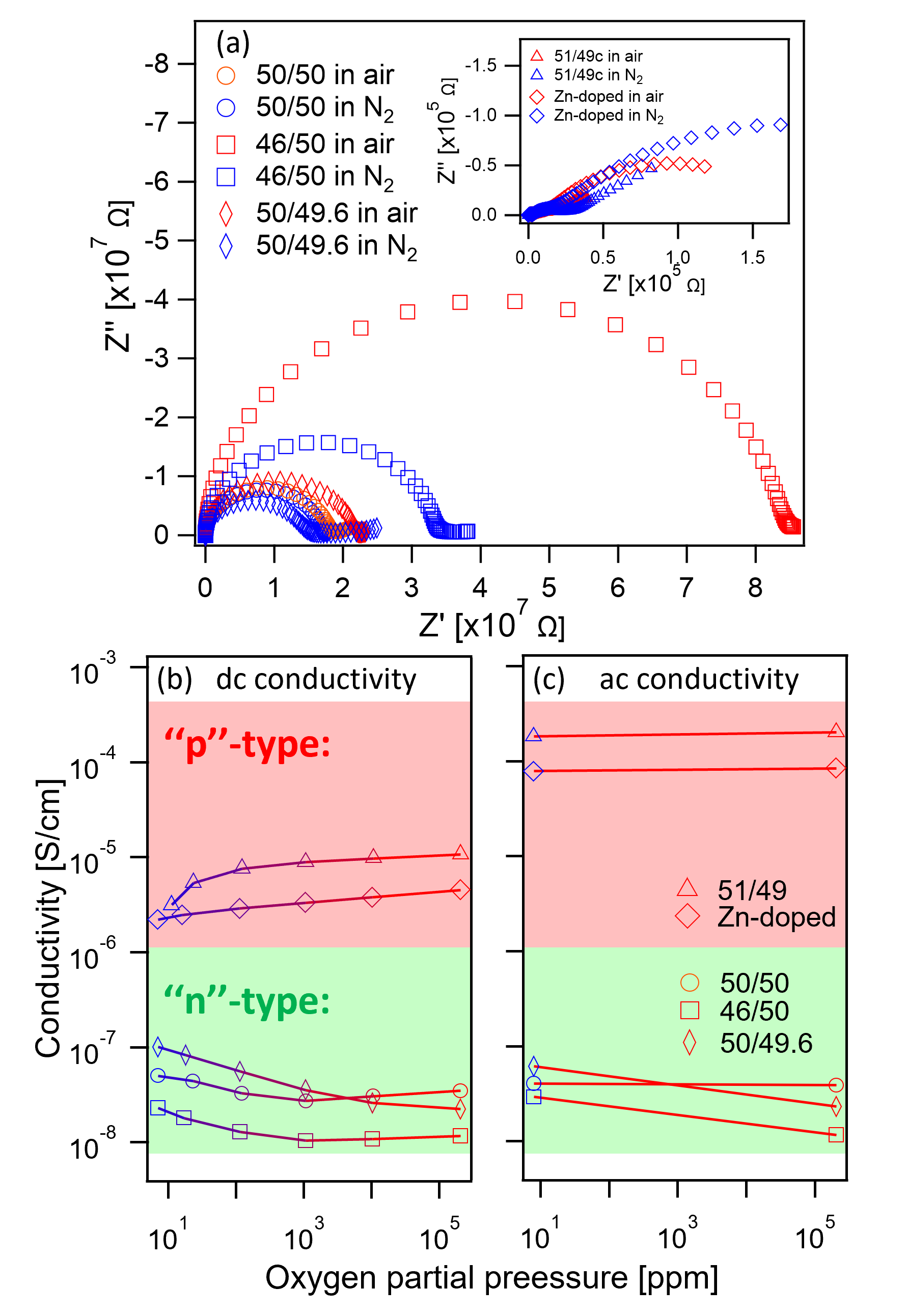}
    \caption{(a) Nyquist plots of impedance measurements performed in air and N$_2$ atmosphere for four different samples at 450$^{\circ}$C. Equilibrium dc and ac conductivities at different oxygen partial pressures recorded at 450$^{\circ}$C are displayed in (b) and (c), respectively. The sample compositions are indicated by different symbols in (a), which are consistently used in (b) and (c). }
    \label{fig 2}
\end{figure}

The oxygen partial pressure dependence is measured by both dc and ac methods. The dc conductivities, shown in Fig. \ref{fig 2} (b), are extracted from current–time measurements, performed at different oxygen partial pressures at a constant temperature (static method) of $450\,^\circ$C. The time-dependent curves are displayed in Fig. \ref{fig A.1.1} in the Supplementary Information. The conductivities at a given partial pressure are extracted from the curves just before the partial pressure is changed again for the next measurement point. The ac conductivity, shown in Fig. \ref{fig 2} (c), is obtained by the bulk contribution of the samples in Nyquist plots shown in Fig. \ref{fig 2} (a). Five representative results for samples 50/50, 51/49, 46/50, 50/49.6 and the Zn-doped sample are displayed here, while results for additional samples are provided in Fig. \ref{fig A.1.2} of the Supplementary Information.

Nyquist plots (Fig. \ref{fig 2} (a)) show that the 50/50, 46/50 and 50/49.6 samples exhibit a single semicircle in dry air or N$_2$, which can be modeled using a single RC circuit, indicating that only the bulk contribution is present in these samples. On the other hand, the 51/49 and the Zn-doped samples exhibit more than one semicircle in their Nyquist plots, indicating the presence of both bulk and grain boundary contributions in these samples. More detailed Nyquist plots of 51/49, 50/49 and Zn-doepd samples are shown in Fig. \ref{fig A6} and the extracted bulk resistivities (R$_{\mathrm{b}}$) and grain boundary resistivities (R$_{\mathrm{GB}}$) are summarized in Table \ref{TableS1}.

Two distinct behaviors can be observed from dc and ac conductivity: the 50/50, 46/50 and 50/49.6 samples exhibit comparable dc and ac conductivity, with a value of approximately $10^{-8}$\,S/cm. The dc conductivity increases as the oxygen partial pressure decreases, typical for n-type behavior. In contrast, the 51/49 and Zn-doped samples display an ac conductivity of $10^{-4}$\,S/cm. This is two orders of magnitude higher than their dc conductivity, which are approximately $5 \times 10^{-6}$\,S/cm. Moreover, the dc conductivity in these samples decreases with decreasing oxygen partial pressure, indicative of p-type behavior.

By comparing the dc conductivity (electronic conductivity) and ac conductivity (total conductivity) at $450\,^{\circ}$C, all samples with varying $A$-site stoichiometry and 1\% Zn doping exhibit two distinct types of behavior:
\begin{itemize}
	\item Type I (``p''-type): observed in Na-rich and acceptor-doped samples, effectively acceptor-doped, and characterized by high ionic conductivity accompanied by p-type electronic conductivity at $450\,^{\circ}$C.
	\item Type II (``n''-type): found in slightly Na-rich and Bi-rich samples, effectively donor-doped, and dominated by less conductive n-type electronic conductivity. Some samples exhibit slight p-type behavior but remain significantly more insulating than Type I samples at $450\,^{\circ}$C.
\end{itemize}

The ``p''-type electronic conduction of the Na-excess samples can be attributed to the presence of $\mathrm{Na_{Bi}^{\prime\prime}}$, as discussed in the outline of the defect model. Those $\mathrm{Na_{Bi}^{\prime\prime}}$ antisite acceptor defects are partially compensated by holes to maintain charge neutrality. Additionally, Zn-doping, presumed to substitute at the $B$-site, is likely to create point defects of $\mathrm{Zn_{Ti}^{\prime\prime}}$. These extrinsic acceptor defects are also partially compensated by holes to fulfill the charge neutrality condition, contributing to the observed ``p''-type behavior. The nominal stoichiometric sample exhibits a background donor effect, leading to ``n''-type electronic conductivity. Although the 50/49.6 sample is Na-excess, it does not generate sufficient $\mathrm{Na_{Bi}^{\prime\prime}}$ to induce ``p''-type behavior (see Fig. \ref{fig A.1.2} in the Supplementary Information). On the other hand, the 54/50 sample is more Na-excess but remains ``n''-type due to the presence of a Na-excess secondary phase (Fig. \ref{fig A.1.2}). This secondary phase apparently maintains the stoichiometry of the main ceramic phase, rendering it similar to that of a stoichiometric sample.

Furthermore, the ``p''-type samples predominantly exhibit ionic conductivity with the value of $10^{-4}$\,S/cm at $450\,^{\circ}$C, as indicated by the total conductivity being one to two orders of magnitude higher than the electronic conductivity. This behavior can be attributed to the partial compensation of $\mathrm{Na_{Bi}^{\prime\prime}}$ in the Na-excess samples by oxygen vacancies ($\mathrm{V_O^{\cdot\cdot}}$):
\begin{equation}
	\mathrm{V_O^{\cdot\cdot} = Na_{Bi}^{\prime\prime}}
\end{equation}
For the Zn-doped sample, the predominant ionic conductivity is attributed to the partial compensation of oxygen vacancies for the point defects of $\mathrm{Zn_{Ti}^{\prime\prime}}$:
\begin{equation}
	\mathrm{ZnO\longrightarrow Zn_{Ti}^{\prime\prime}+V_{O}^{\cdot\cdot}+O_{O}^{x}}
\end{equation}
The total conductivity of ``p''-type samples remains independent of oxygen partial pressure, which is attributed to the high concentration of $\mathrm{V_O^{\cdot\cdot}}$.

By defining these two types of electrical conductivity behaviors, data from the literature can be analyzed. Pure NBT, synthesized by University of Sheffield (Sheffield), consistently exhibits a background acceptor-doping effect and demonstrates Type I behavior \cite{li2014family,yang2018defect,yang2017enhanced,li2015dramatic}. Similarly, NBT-6BT synthesized by the Sheffield group should also exhibit Type I behavior, as BaTiO$_{3}$ serves as an isovalent substitution, making it electrically neutral and unlikely to alter the conductivity characteristics. The conductivity behavior of the Sheffield samples can be altered from Type I to Type II through 0.5\% Nb doping, which has been attributed to the suppression of oxygen vacancies \cite{li2015donor}. While donor doping does reduce the $\mathrm{V_O^{\cdot\cdot}}$, the suppression of oxygen vacancies is not the primary cause of the transition of electronic conductivity, as $\mathrm{V_O^{\cdot\cdot}}$ also act as donor. Instead, the key factor is the shift from background acceptor-doping to donor-doping. Another strategy employed by the Sheffield group to alter conductivity behavior involves using addition of BiAlO$_3$ \cite{yu2008dielectric, yang2017suppression}. The effect of the addition of BiAlO$_3$ on the electrical conductivity has been attributed to an immobilization of oxygen vacancies by forming defect associates with Al. In contrast, we argue that such an immobilization would only suppress ionic conductivity but will not affect p-type electronic conduction. If only ionic conduction were suppressed, the overall reduction in conductivity would be much less pronounced, as the p-type electronic conductivity is still significantly higher than the n-type electronic conductivity in NBT. Therefore, the significant decrease in conductivity by addition of BiAlO$_3$, suggests a shift from effective acceptor-doping to effective donor-doping (for example by inducing Bi excess and formation of Al-rich secondary phases). Additionally, an excess of Bi (Na$_{0.5}$Bi$_{0.51}$TiO$_3$) can also shift the conductivity behavior from Type I to Type II \cite{li2014family}. In summary, the Sheffield samples consistently exhibits a background effective acceptor doping, likely due to Bi loss during sintering, resulting in Type I behavior. This behavior can be transformed into Type II through Bi excess, donor-doping, or solid solution formation with BiAlO$_3$, which all shift the sample from effective acceptor doping to effective donor doping.

In contrast to the samples processed by the Sheffield group, NBT and NBT-6BT samples prepared at Graz University of Technology (TUG) and at Technical University of Darmstadt (TUDa) consistently exhibit a background donor doping, resulting in Type II behavior (this work and \cite{steiner2019effect, seo2017effect, tuprints24767}). This conductivity type can be converted from Type II to Type I through Na excess (Na$_{0.5}$Bi$_{0.49}$TiO$_3$) \cite{seo2017effect} and acceptor doping, such as 1\% Fe in NBT \cite{steiner2019effect}, 2\% Fe in NBT-6BT \cite{steiner2019effect}, and 1\% Zn in NBT-6BT \cite{tuprints24767}. These modifications are believed to shift the sample from effective donor doping to effective acceptor doping.

\subsection{Comparison of experiments with the defect model}

\begin{figure}[ht]
	\centering
	\includegraphics[scale=0.65]{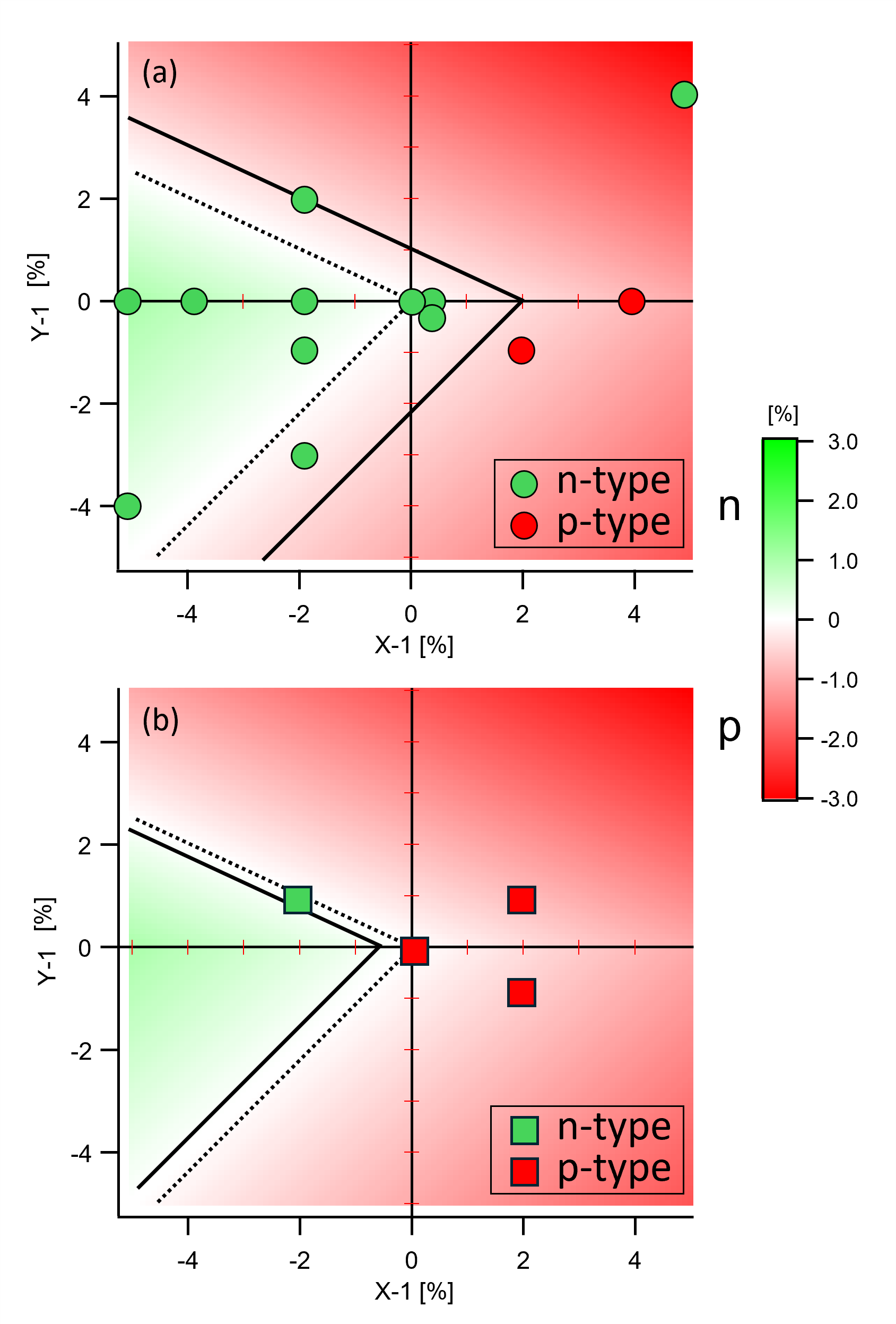}
	\centering
	\caption{Two-dimensional schematic diagrams showing the effective doping concentration as a function of X (Na:Bi ratio) and Y ($A$:$B$ ratio) for (a) samples from this work and (b) samples from the University of Sheffield \cite{yang2018defect}. Green and red regions represent effective donor and acceptor doping, respectively. Green symbols denote ``n''-type samples, while red symbols indicate ``p''-type samples. Dashed and solid lines represent the boundaries between donor and acceptor regions, with and without considering the effect of background doping, respectively.} 
	\label{fig 5.2.1}
\end{figure}

Fig. \ref{fig 5.2.1} (a) and (b) illustrate the two-dimensional schematic diagrams mapping the effective doping concentration as a function of $X$ (Na:Bi ratio) and $Y$ ($A$:$B$ ratio). In this representation, green and red colors indicate effective donor- and acceptor-doping, respectively, with darker shades corresponding to higher doping concentrations. The conduction types of all samples with varying $A$-site stoichiometry from this work are also included in Fig. \ref{fig 5.2.1} (a).

The dashed line represents the calculated boundary between different effective doping regions. As showed in Fig. \ref{fig 2}, our samples consistently exhibit a background donor doping. Since the conduction type can be switched by 1\% Zn doping, which correspond to a 0.02 effective acceptor doping, the boundary line (solid line) is shifted by 0.02 towards the effective acceptor region. This adjustment aligns well with the experimental results and accurately predicts the conduction type. The only exception is the 54/50 sample, which falls within the effective acceptor region but remains ``n''-type. This can be explained by the presence of a Na-excess secondary phase (see Fig. \ref{fig A.1.7}), which effectively removes the Na excess and leaves the majority of grains in their preferred ``n''-type composition.

Next, we apply the defect model to predict the conduction behavior of the samples from the Sheffield group, which consistently exhibit a background acceptor doping. Since the conduction type can be altered by 0.5\% Nb doping—introducing approximately 0.01 effective donor doping, we have shifted the boundary line (solid line) by 0.01 towards the effective donor region. Fig. \ref{fig 5.2.1} (b) demonstrates that the data from the Sheffield group align well with our defect model.

In summary, the defect model, which accounts for both $A$-site Na:Bi and the $A$:$B$ ratio, effectively predicts the conduction mechanism by adjusting the boundary based on the background donor or acceptor doping.

\subsection{Temperature dependence}

\begin{figure}[ht]
    \centering
    \includegraphics[width=8.5cm]{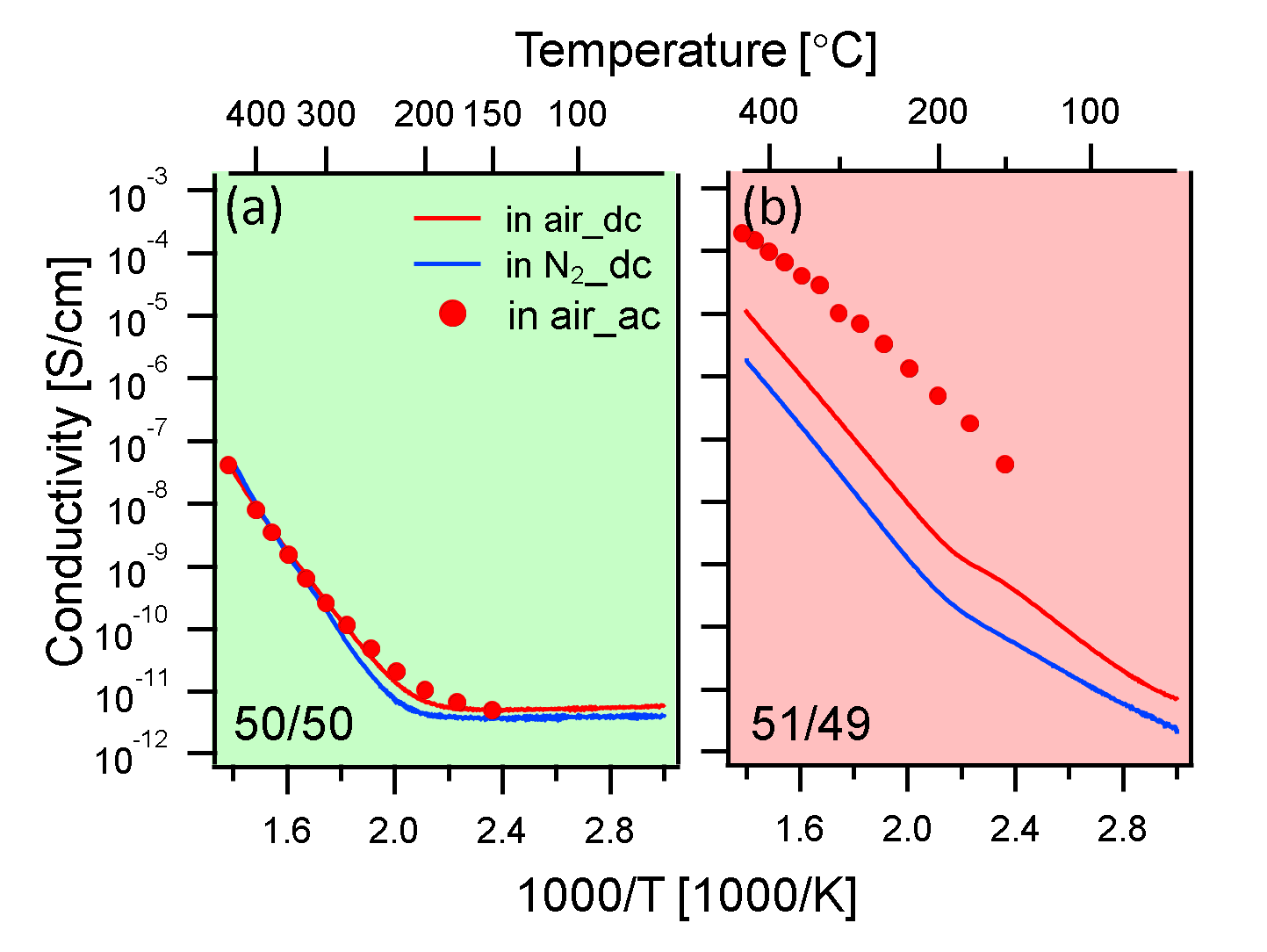}
    \caption{ The Arrhenius plots of conductivity for NBT-6BT ceramics of (a) 50/50 and (b) 51/49, measured in dry air, and the dynamic dc method (solid lines in different colors) with a voltage of 0.5\,V. The red lines represent measurements in dry air, while other colors correspond to measurements in N$_2$, plotted as a function of 1000/T.}
    \label{fig 3}
\end{figure}

Fig. \ref{fig 3} presents the Arrhenius plots of bulk conductivity for two representative samples (50/50 and 51/49), measured using ac static method (solid symbols) and dc dynamic method (solid lines) in dry air and N$_2$, corresponding to oxygen partial pressures of 10$^5$ and 10$^0$ ppm, respectively. It is noted that only the final cooling cycle is displayed, as the initial heating and cooling cycles primarily reflect an equilibration process. Additional data for other samples are provided in Fig. \ref{fig A.1.3} in the Supplementary Information. Corresponding Nyquist plots for these samples, measured over the temperature range of 150–450$^{\circ}$C, are also included in Fig. \ref{fig A.1.4} in the Supplementary Information.

In dry air, the electronic conductivity behavior of the 50/50 sample exhibit two distinct regions. At lower temperatures (below approximately $200\,^{\circ}$C), a plateau is observed, which can be attributed to the charging and discharging currents of the capacitor and to the lower current detection limit of the picoammeter, rather than to a lower activation energy. While at higher temperatures (above approximately $200\,^{\circ}$C), the conductivity increases exponentially with reciprocal temperature, indicating thermally activated behavior. In contrast, the 51/49 sample displays two clearly separated linear regions with distinct activation energies, divided around $200\,^{\circ}$C. In N$_2$, the electronic conductivity behavior of all samples follows similar trends to those observed in dry air.

The conduction type can be further verified by comparing the ac and dc conductivity at $450\,^{\circ}$C under dry air and N$_2$ atmospheres. These results are consistent with the oxygen partial pressure–dependent measurements, confirming that only the 51/49, 50/49, and 1\% Zn-doped samples exhibit dominant ionic conductivity ($\sigma_{\mathrm{ac}}\gg\sigma_{\mathrm{dc}}$), whereas all other samples exhibit dominant electronic conductivity ($\sigma_{\mathrm{ac}}\approx\sigma_{\mathrm{dc}}$). All Type II samples display an electronic conductivity of approximately $10^{-8}\,$S/cm at $450\,^{\circ}$C in dry air, whereas the Type I samples demonstrate p-type electronic conductivity that is 2–3 orders of magnitude higher.

\begin{figure}[ht]
    \centering
    \includegraphics[width=8.5cm]{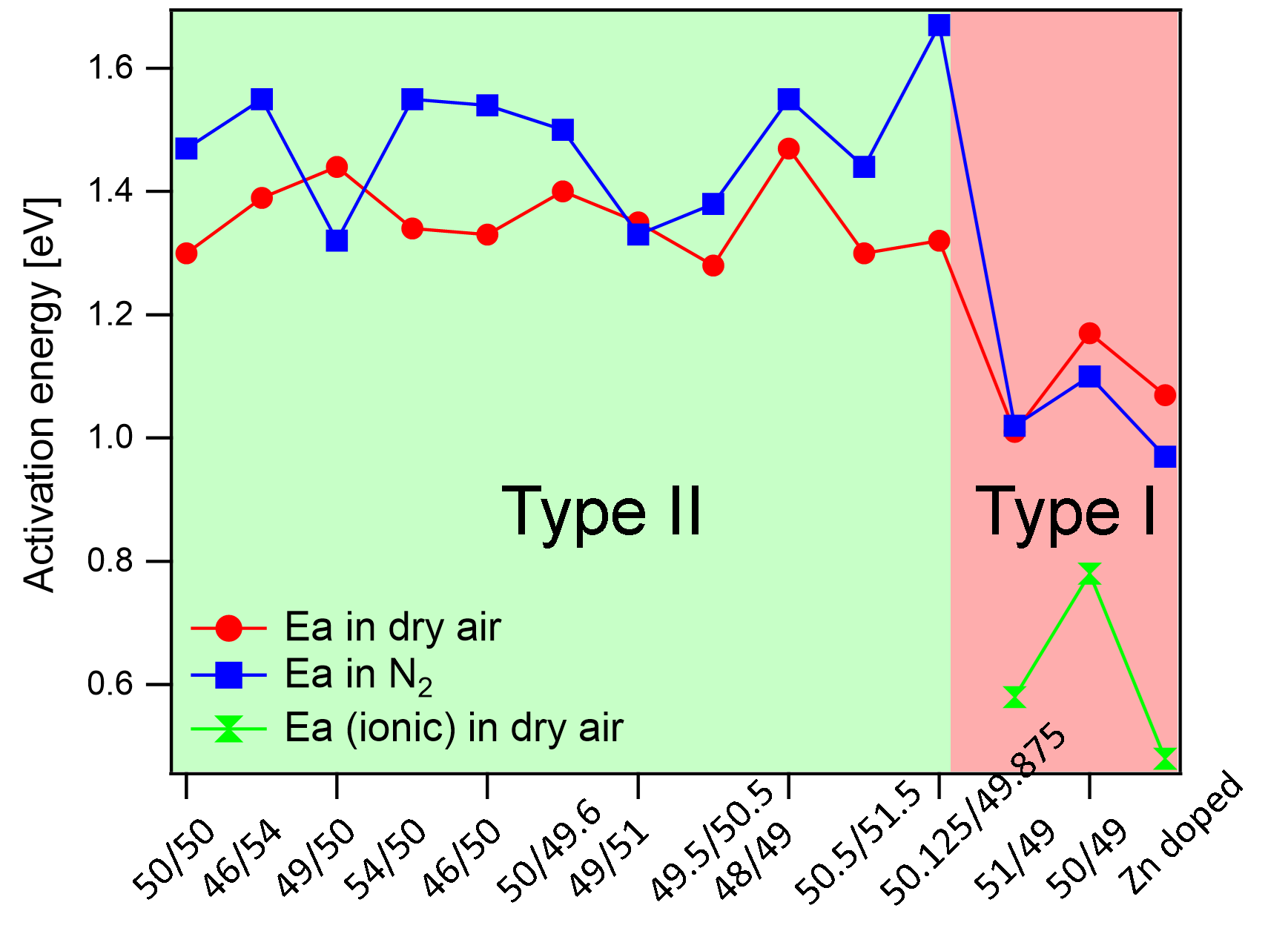}
    \caption{The activation energies, derived from the dynamic dc method Arrhenius plots between $400\,^\circ$C and $450\,^\circ$C for NBT-6BT ceramics with varying $A$-site nonstoichiometries and $A$:$B$ ratios. The red and blue symbols represent the different atmospheres: dry air and N$_2$ corresponding to oxygen partial pressures of $10^5$ and $1$\,ppm, respectively.}
    \label{fig 4}
\end{figure}

The activation energies, which are shown in Fig. \ref{fig 4}, were taken from the dynamic dc method in dry air and N$_2$ between $400\,^\circ$C and $450\,^\circ$C, according to the Arrhenius relation:

\begin{equation}
	\sigma_{\mathrm{dc}}=\sigma_{0}\exp\left(-\frac{E_{\mathrm{a}}}{k_{\mathrm{B}}\mathrm{T}}\right)
\end{equation}

where $\sigma_0$ is pre-exponential factor (related to carrier mobility and concentration), $k_{B}$ is Boltzmann constant ($8.617 \times 10^{-5}\,$eV/K) and $T$ is absolute temperature (in Kelvin).

\begin{itemize}
	\item Type I samples: exhibit lower electronic $E_{\mathrm{a}}$ in both dry air and N$_2$ compared to the Type II samples. Specifically, the electronic $E{_{\mathrm{a}}}$ values in dry air range from approximately 1.0 to 1.16\,eV, which are slightly lower than their $E_{\mathrm{a}}$ values in N$_2$, also falling within this range except for the 50/49 sample. Since the total conductivity is dominated by ionic transport, the activation energies extracted from total conductivity are considered representative of ionic conductivity, ranging from 0.5 to 0.7\,eV. These values align well with reported ionic activation energies in the literature \cite{li2014family,li2015dramatic,yang2018defect,yang2017enhanced}, corresponding to oxygen vacancy migration energies. However, there is limited literature available on the activation energy of electronic conductivity in Type I samples for direct comparison.
	
	\item Type II samples: exhibit higher electronic $E_{\mathrm{a}}$ in both dry air and N$_2$ compared to the Type I samples. Specifically, the electronic activation energy ($E_{\mathrm{a}}$) values in N$_2$ range from approximately 1.4 to 1.6\,eV, which are higher than their corresponding values in dry air, except for the 49/50 and 49/51 samples, where the values are slightly lower (1.3 to 1.4\,eV). The electronic $E{_{\mathrm{a}}}$ values measured in air are consistent with those reported in the literature for Type II samples from TUDa, which fall within the range of 1.3 to 1.4\,eV \cite{seo2017effect,steiner2019effect}. However, these values are lower than those observed in Bi-excess Type II samples from Sheffield, which exhibit an electronic $E_{\mathrm{a}}$ of approximately 1.7\,eV \cite{li2015dramatic}.
\end{itemize}

\subsection{$A$-site nonstoichiometry and Fermi level in NBT-6BT ceramics}

\begin{figure*}[tbp]
	\centering
	\includegraphics[scale=0.8]{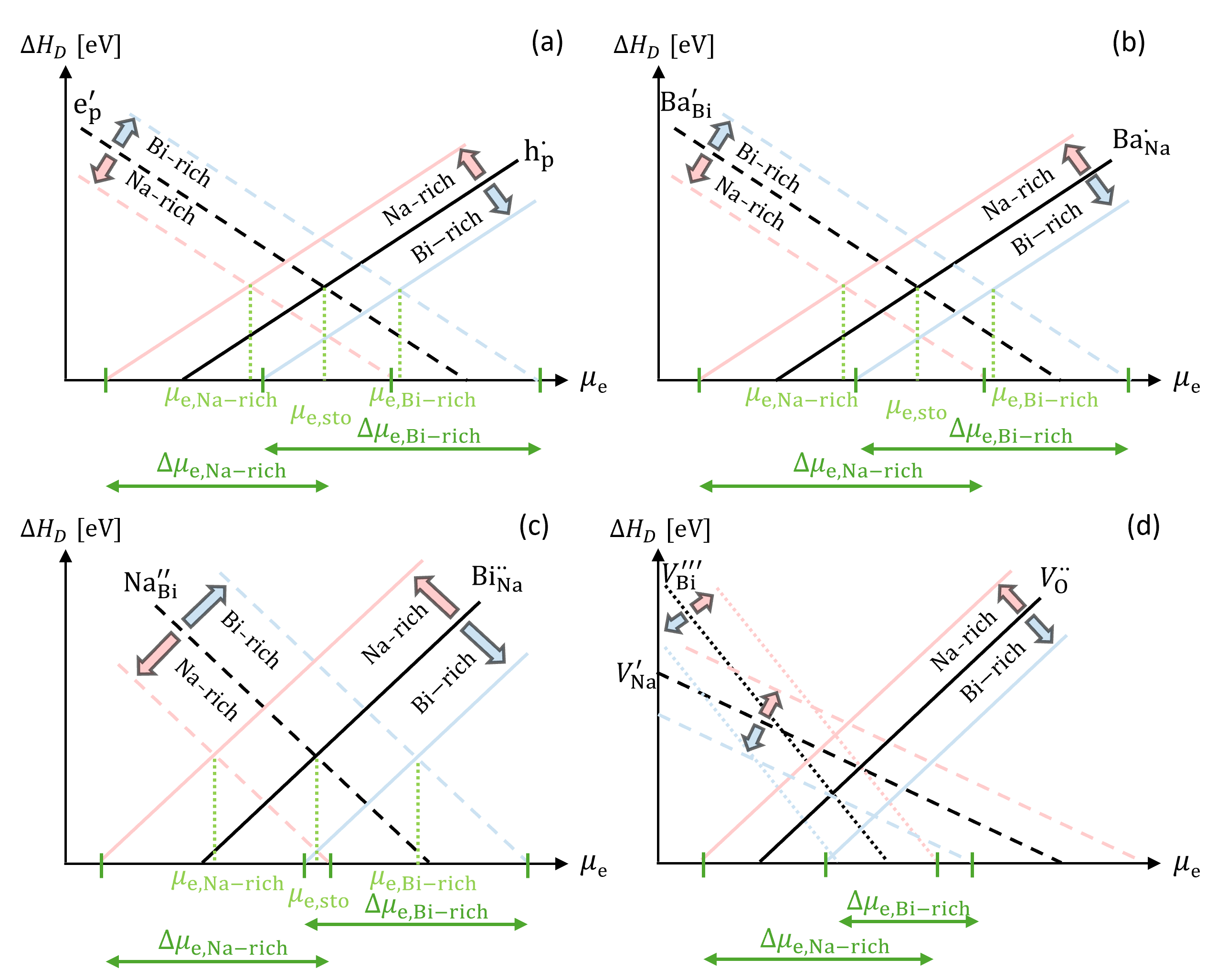}
	\centering
	\caption{The formation enthalpies of (a) $\mathrm{e_{p}^{\prime}}$ (dashed lines) and $\mathrm{h_{p}^{\cdot}}$ (solid lines), (b) $\mathrm{Ba_{Bi}^{\prime}}$ (dashed lines) and $\mathrm{Ba_{Na}^{\cdot}}$ (solid lines), (c) $\mathrm{Na_{Bi}^{\prime\prime}}$ (dashed lines) and $\mathrm{Bi_{Na}^{\cdot\cdot}}$ (solid lines) and (d) $\mathrm{V_{Na}^{\prime}}$ (long dashed lines), $\mathrm{V_{Bi}^{\prime\prime\prime}}$ (short dashed lines) and $\mathrm{V_{O}^{\cdot\cdot}}$ (solid lines) as a function of the Fermi level position ($\mu_{\mathrm{e}}$) in NBT-6BT. The red and blue arrows indicate the shifts in defect formation enthalpies under Na-rich and Bi-rich compositions, respectively. The dark green arrows represent the range of Fermi level values over which the material remains thermodynamically stable under these compositions. The light vertical green lines mark the Fermi level positions at which two defects exhibit equal formation enthalpies.} 
	\label{fig 6}
\end{figure*}

Fig. \ref{fig 6} schematically represents the formation enthalpies of various point defects in NBT-6BT as a function of the Fermi level ($\mu_{\mathrm{e}}$): (a) $\mathrm{e_{p}^{\prime}}$ (dashed lines) and $\mathrm{h_{p}^{\cdot}}$ (solid lines), (b) $\mathrm{Ba_{Bi}^{\prime}}$ (dashed lines) and $\mathrm{Ba_{Na}^{\cdot}}$ (solid lines), (c) $\mathrm{Na_{Bi}^{\prime\prime}}$ (dashed lines) and $\mathrm{Bi_{Na}^{\cdot\cdot}}$ (solid lines) and (d) $\mathrm{V_{Na}^{\prime}}$ (long dashed lines), $\mathrm{V_{Bi}^{\prime\prime\prime}}$ (short dashed lines) and $\mathrm{V_{O}^{\cdot\cdot}}$ (solid lines). This figure serves as a qualitative illustration of defect formation enthalpies under Na-rich and Bi-rich compositions. It does not include charge transition levels, and the intersection points where two defects exhibit equal formation enthalpies are not quantitative. In particular, Fig. \ref{fig 6} (d) includes three types of vacancies, resulting in a complex defect landscape with limited quantitative reliability.

Bi-rich and Na-rich compositions can be regarded as exhibiting effective donor-like and acceptor-like behavior, respectively, referred to as ``self-doping'' in this work. Although these compositions do not change the nominal valence states of Bi or Na, they can promote the formation of antisite defects (such as $\mathrm{Na_{Bi}^{\prime\prime}}$ and $\mathrm{Bi_{Na}^{\cdot\cdot}}$) or $A$-/$B$-site vacancies, depending on the $A$:$B$ stoichiometric ratio, as discussed in Section of defect model. These intrinsic donor- or acceptor-like effects are always accompanied by a shift in the Fermi level.

Under Bi-rich compositions, which act as effective donor doping, the system donates electrons to the lattice, resulting in positively charged donor species and an upward shift in the Fermi level. This is reflected in Fig. \ref{fig 6}, where all defect formation enthalpy lines shift toward higher Fermi levels. Conversely, Na-rich compositions function as effective acceptor doping, leading to a downward shift in the Fermi energy, as indicated by the corresponding leftward shift in the formation enthalpy lines in Fig. \ref{fig 6}. These Fermi level shifts significantly influence the formation enthalpies of various defects, which could be involved in the compensation of the charge of the dopant. An upward Fermi level shift (Bi-rich) reduces the formation enthalpy of: i) the trapping of electrons on cation sites, indicated in Fig. \ref{fig 6} (a), ii) $\mathrm{Ba_{Bi}^{\prime}}$ substitution, indicated in Figure \ref{fig 6} (b), iii) $\mathrm{Na_{Bi}^{\prime\prime}}$ antisite defects, indicated in Fig. \ref{fig 6} (c), and iv) Bi and Na vacancies, indicated in Fig. \ref{fig 6} (d). Conversely, a downward Fermi level shift (Na-rich) has the same effects of the formation enthalpy for trapped holes, $\mathrm{Ba_{Na}^{\cdot}}$ substitution, $\mathrm{Bi_{Na}^{\cdot\cdot}}$ antisite defects, and oxygen vacancies.

It is noted that, according to Fig. \ref{fig 6} (d), at higher Fermi level s, the formation of $\mathrm{V_{Bi}^{\prime\prime\prime}}$ is generally expected to be more favorable than that of $\mathrm{V_{Na}^{\prime}}$. Under oxidizing conditions, i.e.\ at low Fermi levels, $\mathrm{V_{Na}^{\prime}}$ become the dominant defect species. 

Furthermore, the limits of the Fermi level are defined by the points at which defect formation enthalpies approach zero, marking the onset of spontaneous defect formation. As shown in Fig. \ref{fig 6}, the double-sided arrows with dark green color indicate the Fermi level range for each pair of defects within which the material remains thermodynamically stable under Na-rich and Bi-rich compositions. However, the precise determination of the overall stability range of the Fermi level under these chemical potential conditions remains elusive, as we are unable to combine all relevant defect formation enthalpies into a single, comprehensive plot. This limitation arises from the lack of reliable defect formation energy data derived from density functional theory (DFT) calculations. Computing such energies in NBT and its solid solution NBT-BT, where Ba substitutes on the $A$-site, is particularly challenging due to the intrinsic $A$-site disorder caused by the random distribution of Na and Bi atoms in the real material. In contrast, DFT simulations require a predefined $A$-site ordering, which has been shown to significantly influence the calculated defect formation energies \cite{tuprints19898}. Importantly, previous in situ XPS experiments using an electrochemical-cell setup demonstrated the emergence of metallic Bi when the Fermi level reached its upper limit—corresponding to the right boundary in Fig. \ref{fig 6}, located approximately 2.47\,eV above the valence-band minimum \cite{hu2024fermi}, thereby experimentally confirming this stability threshold.

\section{Conclusion}

A new defect model is proposed to elucidate the combined effects of $A$-site non-stoichiometry and $A$:$B$ ratios in NBT-BT ceramics. Electrical conductivity measurements demonstrate that Na-excess samples exhibit dominant ionic conductivity and ``p''-type behavior, while Bi-excess samples display insulating ``n''-type electronic conductivity, strongly validating the model. This new defect model offers a more comprehensive understanding of the defect chemistry in NBT-BT ceramics and serves as a valuable guide for optimizing sample processing to achieve tailored properties in NBT-BT materials.

\section*{Acknowledgement}

The presented work has been conducted within the collaborative research centre FLAIR (Fermi level engineering applied to oxide electroceramics), supported by the German Research Foundation (DFG) (project-ID 463184206 -- SFB 1548) and by the Austrian Fonds zur F{\"{o}}rderung der wissenschaftlichen Forschung (FWF, Project Grant-DOI (10.55776/I6450)). Pengcheng Hu also acknowledges support from the China Scholarship Council (CSC) (Award No. 202106220039). For the purpose of open access, the author has applied a CC BY public copyright license to any Author Accepted Manuscript version arising from this submission. 

\bibliographystyle{elsarticle-num}
\bibliography{NBTnp}

\newpage

\onecolumngrid
\appendix
\section*{Supplementary Information}

\begin{figure}[h!]
	\centering
	\includegraphics[scale=1.2]{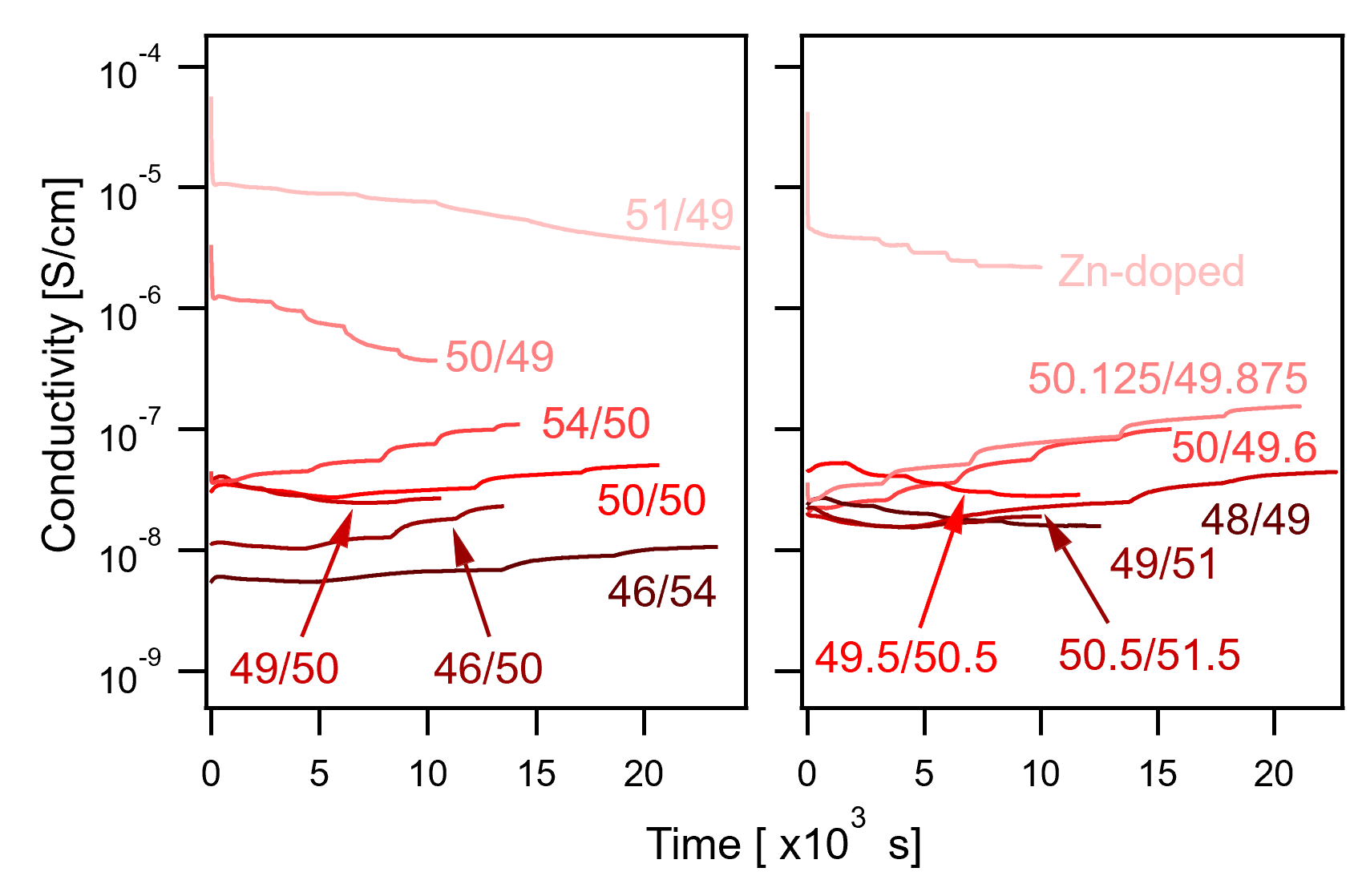}
	\centering
	\caption{The conductivity-time plots measured at 450$^{\circ}$C using a dc voltage of 0.5V over an oxygen partial pressure range from 10$^5$ ppm to 10$^{0}$ ppm on NBT-6BT ceramics with different A-site stoichiometry and 1\% Zn doping. The red lines represent measurements conducted over oxygen partial pressure ranges of 10$^5$ to 10$^0$ ppm.} 
	\label{fig A.1.1}
\end{figure}

Fig. \ref{fig A.1.1} presents the conductivity-time plots from oxygen partial pressure dependence measurements performed at a constant dc voltage of 0.5 V. The measurements span a wide range of oxygen partial pressures, from 10$^5$ to 10$^0$ ppm. These experiments were conducted at 450$^{\circ}$C on various 0.94(Na$_{0.5}$Bi$_{0.5}$)TiO$_3$-0.06BaTiO$_3$ (NBT-6BT) ceramics with different A-site stoichiometry and 1\% Zn doping. The red lines represent the measurements in the oxygen partial pressure ranges of 10$^5$ to 10$^0$ ppm. 

\begin{figure*}[h!]
	\centering
	\includegraphics[scale=0.6]{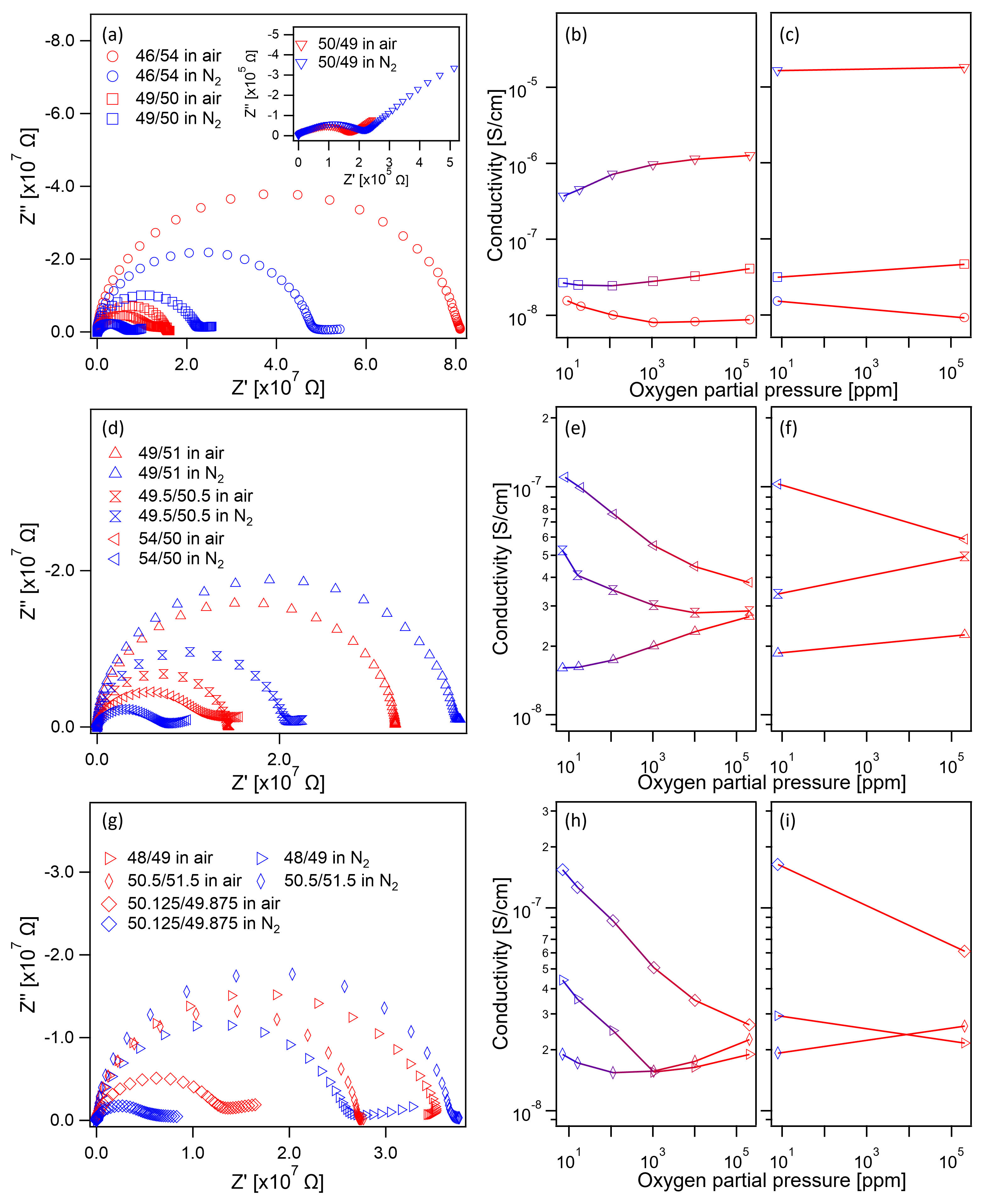}
	\centering
	\caption{(a), (d) and (g): Nyquist plots measured in dry air and N$_2$, and the oxygen partial pressure dependence of (b), (e) and (h): dc conductivity and (c), (f) and (i): ac conductivity for NBT-6BT-based ceramics with varying A-site stoichiometry at 450$^{\circ}$C. The sample compositions are indicated by different symbols in (a), (b), and (c); (d), (e), and (f); and (g), (h), and (i), respectively, with consistent symbol usage within each group.} 
	\label{fig A.1.2}
\end{figure*}

Fig. \ref{fig A.1.2} shows (a), (d), and (g): Nyquist plots in air and N$_{2}$; (b), (e), and (h): the oxygen partial pressure dependence of dc conductivity; and (c), (f), and (i): the oxygen partial pressure dependence of ac conductivity for NBT-6BT ceramics with varying A-site stoichiometry at 450 $^{\circ}$C. The sample compositions are indicated by different symbols in (a)-(c); (d)-(f); and (g)-(i), respectively, with consistent symbol usage within each group. The plateau observed below approximately 200$^{\circ}$C is attributed to the charging and discharging currents of the capacitor, as well as the lower current detection limit of the picoammeter.

\begin{figure*}[h!]
	\centering
	\includegraphics[scale=0.85]{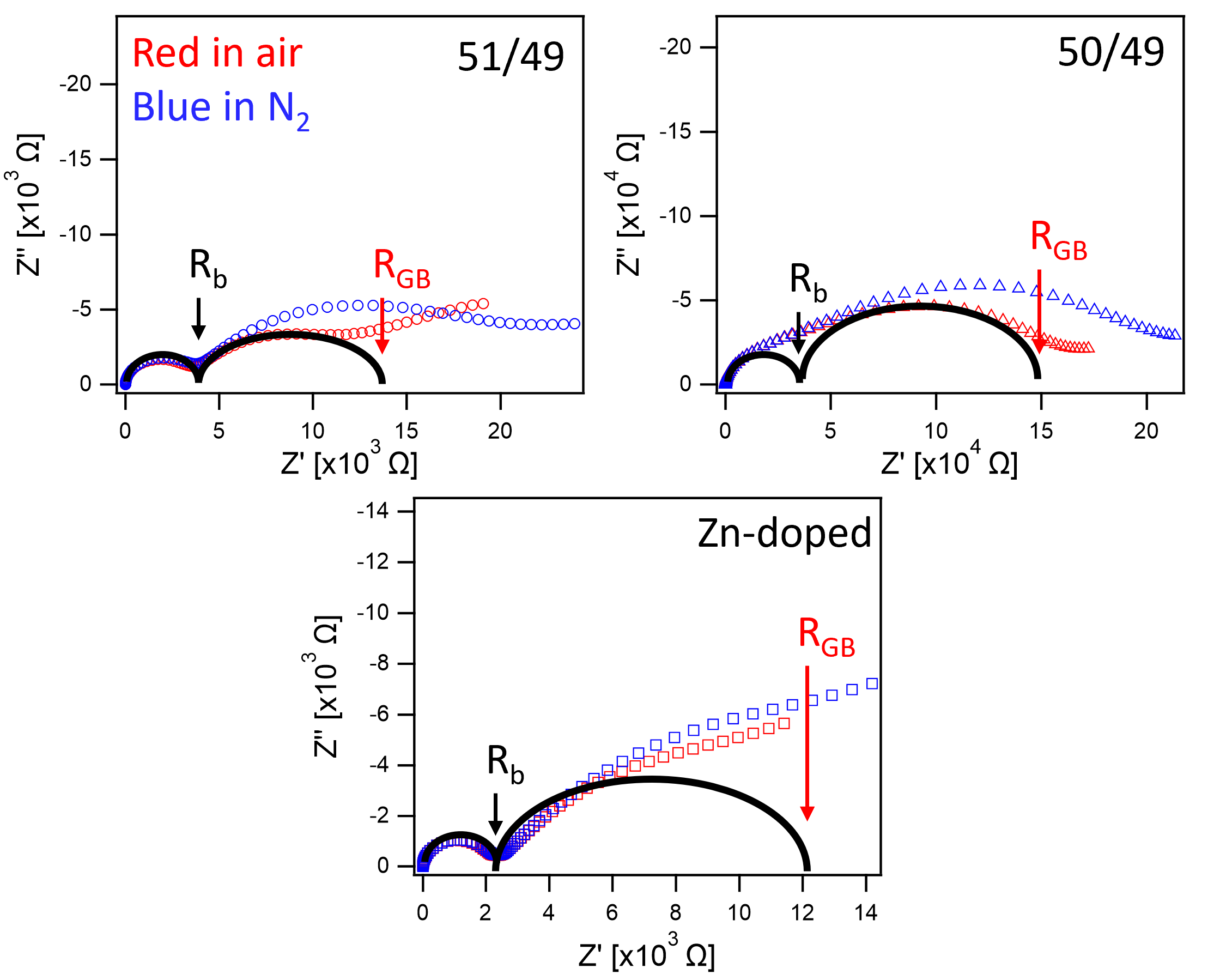}
	\centering
	\caption{Nyquist plots of ``p''-type samples: 51/49, 50/49 and Zn-doepd measured in dry air and N$_2$.} 
	\label{fig A6}
\end{figure*}

Fig. \ref{fig A6} presents the Nyquist plots measured in dry air and N$_2$ for the ``p''-type samples: 51/49, 50/49, and the Zn-doped sample. The extracted bulk resistivity (R$_{\mathrm{b}}$) and grain boundary resistivity (R$_{\mathrm{GB}}$) are summarized in Table \ref{TableS1}. The R$_{\mathrm{b}}$ values were used to determine the ac conductivity, and all samples exhibit higher R$_{\mathrm{GB}}$ in N$_2$ compared to dry air.

\begin{table}[h!]
	\centering
	\caption{Extracted bulk resistivity, R$_\mathrm{b}$, and grain boundary resistivity, R$_\mathrm{GB}$, from  Nyquist plots, Fig. \ref{fig A6}.}
	\label{TableS1}
	\vspace{5pt}
	\small
	\setlength{\tabcolsep}{2.5mm}{
		\begin{tabular}{lcccccc}
			\hline \hline
              & 51/49-air & 51/49-N$_2$ & 50/49-air& 50/49-N$_2$& Zn-doped-air& Zn-doped-N$_2$\\ \hline 
			R$_{\mathrm{B}}$ [$\Omega$] &3.720 $\times$ 10$^{3}$ &1.420 $\times$ 10$^{4}$ &3.953 $\times$ 10$^{4}$& 1.498 $\times$ 10$^{5}$ &2.381 $\times$ 10$^{3}$   &1.224 $\times$ 10$^{4}$ \\
			R$_{\mathrm{GB}}$ [$\Omega$] &4.110 $\times$ 10$^{3}$ &2.080 $\times$ 10$^{4}$ &4.316 $\times$ 10$^{4}$  &2.198 $\times$ 10$^{5}$ & 2.510 $\times$ 10$^{3}$ & 1.824 $\times$ 10$^{4}$\\ \hline 
			\end{tabular}}	
\end{table} 

\begin{figure*}[h!]
	\centering
	\includegraphics[scale=1.5]{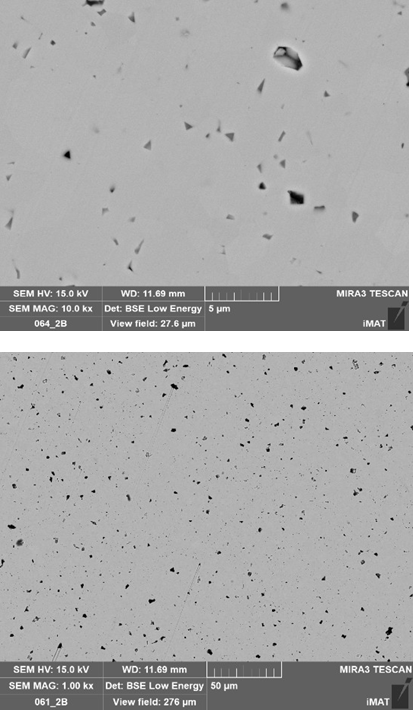}
	\centering
	\caption{Scanning electron microscopic images of 54/50c sample.} 
	\label{fig A.1.7}
\end{figure*}

Fig. \ref{fig A.1.7} presents the scanning electron microscopy images of the 54/50c sample, revealing the presence of a Na-rich secondary phase.

\begin{figure*}[h!]
	\centering
	\includegraphics[scale=0.75]{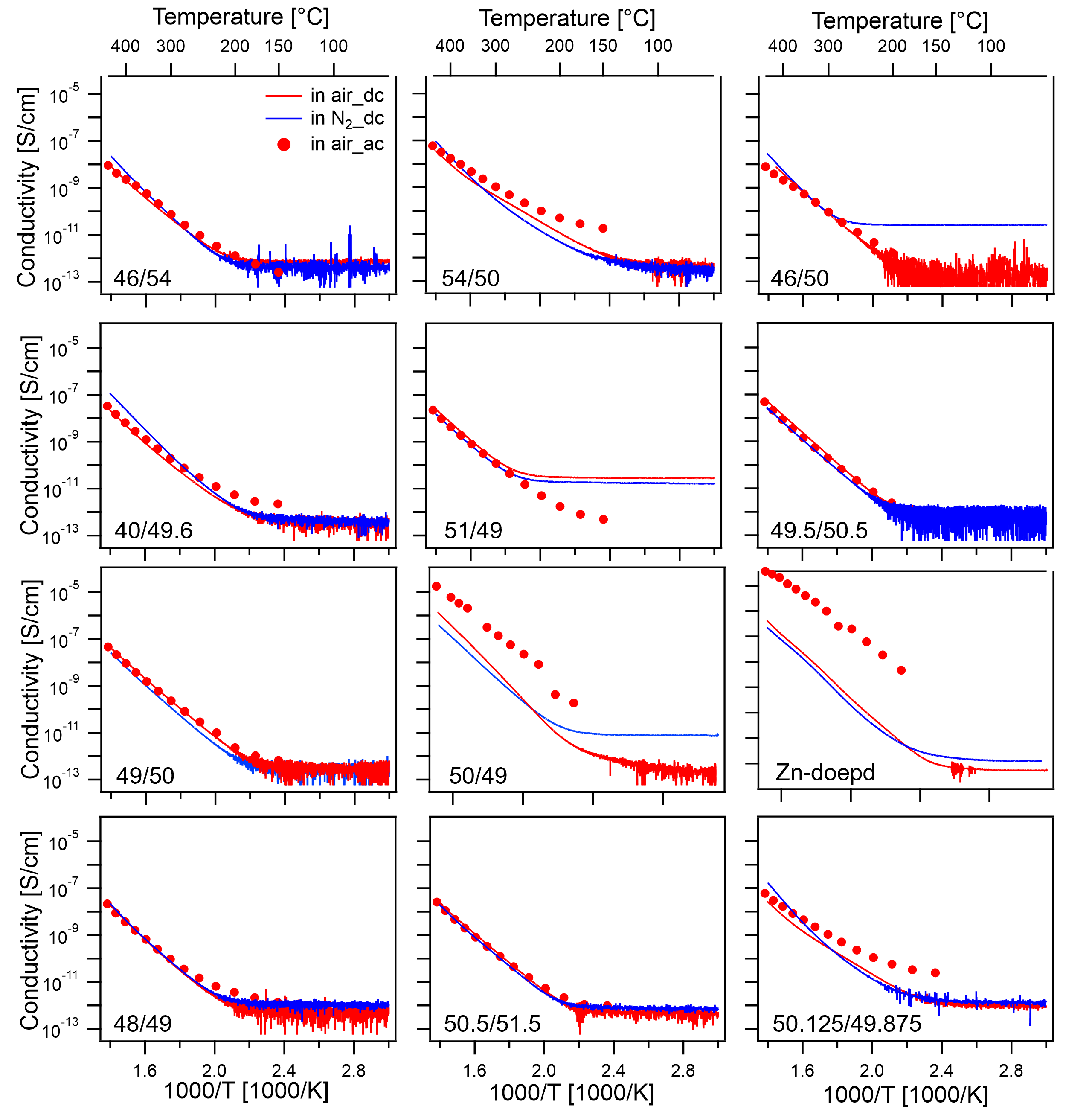}
	\centering
	\caption{The Arrhenius plots of conductivity at ac (solid circles) and dc (solid lines) voltages of 0.5 V for NBT-6BT with various A-site stoichiometry. The red and blue lines represent the different atmospheres used during the measurements: dry air and N$_2$, corresponding to oxygen partial pressures of 10$^5$ and 10$^0$ ppm, respectively. The sample compositions are indicated in the lower left corner of each plot.} 
	\label{fig A.1.3}
\end{figure*}

Fig. \ref{fig A.1.3} shows the Arrhenius plots of bulk conductivity at ac (solid symbols) and dc (solid lines) voltages of 0.5 V of NBT-6BT with various A-site stoichiometry. The sample compositions are indicated in the lower left corner of each plot.

\begin{figure*}[h!]
	\centering
	\includegraphics[scale=0.7]{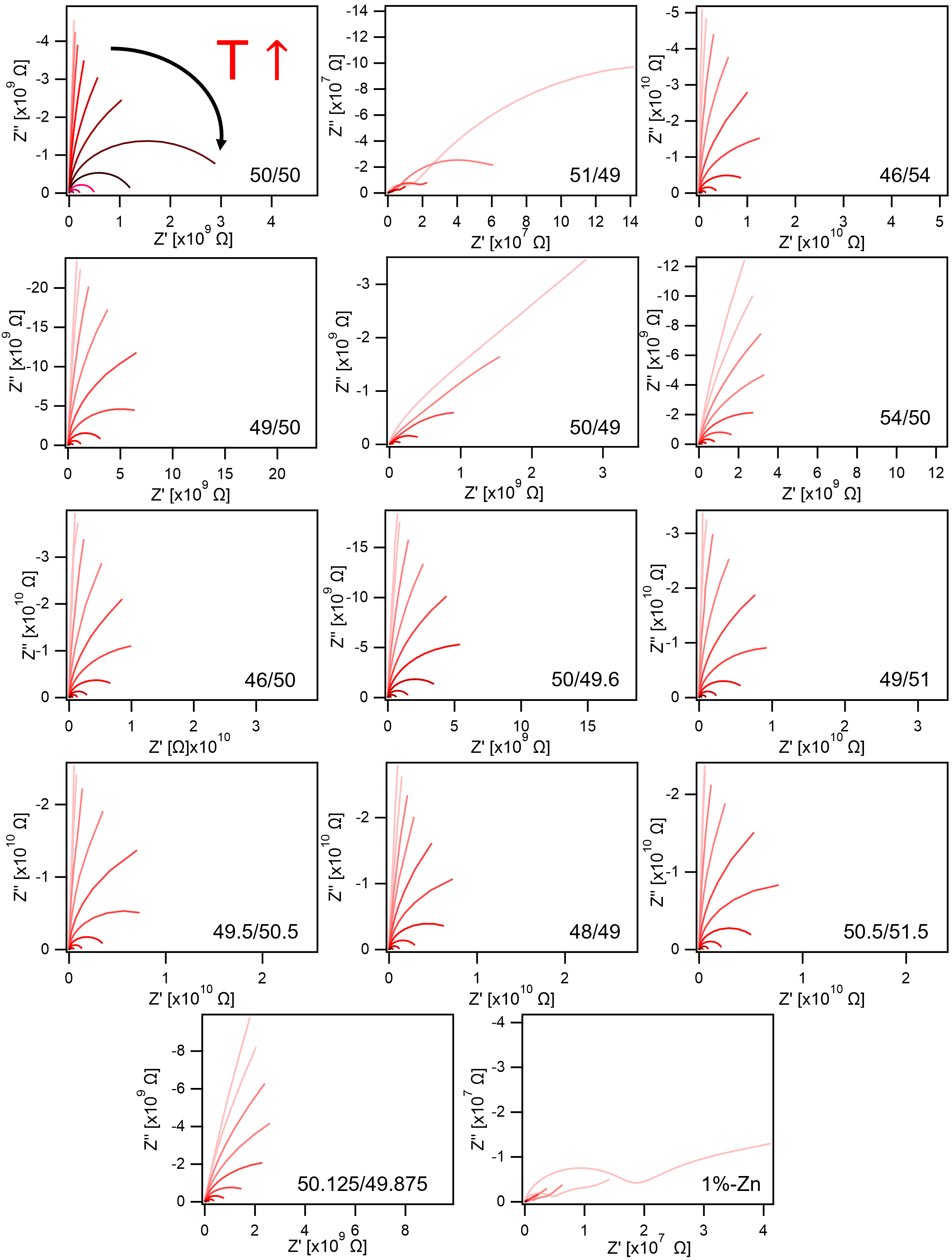}
	\centering
	\caption{The Nyquist plots for NBT-6BT ceramics with various A-site stoichiometry and 1\% Zn doping for temperatures ranging from 150 to 450$^{\circ}$C in increments of 25$^{\circ}$C. The increasing darkness of the line color corresponds to a rise in temperature. The sample compositions are indicated in the lower right corner of each plot.} 
	\label{fig A.1.4}
\end{figure*}

Fig. \ref{fig A.1.4} shows the Nyquist plots of NBT-6BT with various A-site stoichiometry and 1\% Zn doping at temperatures ranging from 150 to 450$^{\circ}$C.

\end{document}